\documentclass[%
 reprint,
 amsmath,amssymb,
 aps,
prb,
]{revtex4-2}

\usepackage{graphicx}

\usepackage{float}
\usepackage{footnote}
\usepackage{placeins}
\usepackage{xcolor}
\usepackage{lipsum}
\usepackage{amsmath}
\usepackage{nccmath}
\usepackage{gensymb}
\usepackage{empheq} 
\usepackage{physics}
\usepackage{dcolumn}
\usepackage{bm}
\usepackage{hyperref}
\usepackage{times}

\newcommand{\bst}{(Bi$_{1-x}$Sb$_x$)$_2$Te$_3$ }

\begin{document}

\preprint{APS/123-QED}

\title{Gate electrode-induced nonreciprocal resistance in topological insulators}

\author{Sofie K\"olling}
\author{Florian R. Westerhof}
\author{Alexander Brinkman}%
\affiliation{%
 MESA+ Institute for Nanotechnology, University of Twente, Enschede, The Netherlands
}%

\date{\today}

\begin{abstract}
A common method of controlling the chemical potential in topological insulators is applying a gate electrode. 
Simultaneously applying high source-drain bias currents can lead to parasitic effects in such devices. 
We derive that these parasitic effects lead to a gradient in the Hall effect along the current lead of a Hall bar.
Consequently, nonreciprocal effects in both longitudinal and Hall voltages appear upon reversing the bias.
These effects scale similarly to the magnetochiral anisotropy, requiring detailed analysis to make a distinction.
Experimentally we show that nonreciprocal effects can appear in materials where magnetochiral anisotropy is not expected while a top gate is present. Without gate electrode, this nonreciprocal effect is found to be absent.
These results show the importance of considering and, if possible, excluding gate electrode-induced effects when searching for nonreciprocal resistance intrinsic to a material.

\end{abstract}

\maketitle
\section{Introduction}

Nonreciprocal resistance has been receiving attention as a promising probe of asymmetric electronic properties of topological insulators. Examples include studying scattering properties between topological surface states and magnetic scattering centers \cite{ye2022nonreciprocal} or between bulk and edge states \cite{yasuda2020large}. 
When both time-reversal and inversion symmetry are broken, the resistance of a material obtains a nonreciprocal term: the transport of right-movers and left-movers is no longer equivalent. The resistance as function of magnetic field ($B$) and current ($I$) is accordingly described by \cite{rikken2001electrical}
\begin{equation}\label{eq:nonrec0}
    R = R_0 (1 + \beta B^2  + \gamma B I),
\end{equation}
where $\beta$ corresponds to the conventional magnetoresistance, and $\gamma$ to the nonreciprocal, nonlinear resistance, also called the magnetochiral anisotropy \cite{tokura2018nonreciprocal}.
Because the nonreciprocal resistance is directly related to the broken symmetries in the system, it serves as a probe for asymmetries in the band structure and scattering processes \cite{nagaosa2024nonreciprocal}.

An exemplary group of materials showing nonreciprocal resistance are magnetic topological insulators. In this group, inversion symmetry can for example be broken by alternating (non)magnetic layers \cite{ye2022nonreciprocal} or by the edge of the sample itself \cite{yasuda2020large}. Time-reversal symmetry is broken by the magnetism.
In such experiments, materials are often tuned from bulk to surface or edge conduction by applying a gate voltage \cite{chang2013experimental}, capacitively tuning the charge on top and bottom surface of the topological insulator \cite{fatemi2014electrostatic}. 
However, regardless of gate tunability, in general the nonreciprocal resistance remains a small effect \cite{tokura2018nonreciprocal}. One aspect not often discussed explicitly is the magnitude of bias current required to probe nonreciprocal effects due to this smallness.

Applying a DC bias can have multiple effects in transport. Several experiments have observed that a finite bias current causes Joule heating, increasing the electron temperature \cite{nandi2018logarithmic, anderson1979possible, abrahams1980non}, however, this effect is symmetric in bias and would not cause nonreciprocity. Another effect, not symmetric in bias, is the altered potential difference between gate and topological surface upon applying a DC bias current, which influences the surface charge carrier density. Due to the applied DC current, a DC voltage develops. When this voltage becomes comparable to typical top gate voltages required to tune the transport properties, a significant effect on transport properties is expected. 
Specifically, we expect that even without tuning the gate voltage, the addition of a metallic gate electrode itself could be sufficient to observe such self-gating effects upon varying the bias current.
Previous experiments in tunnel junctions have speculated on an interplay between bias and gate voltage, but this conclusion has not been solidified \cite{lee2015mapping}.

In the present research, we study the interplay between bias and gate voltage in a set of magnetically doped topological insulator V$_y$(Bi$_{1-x}$Sb$_{x}$)$_{2-y}$Te$_3$ (VBST) Hall bar devices. We systematically limit the number of explanations by studying magnetic/nonmagnetic ($y=0$) devices with/without top gate. Theoretically, we discuss a first order electrostatic model to interpret the data of top-gated devices. Combining the results, we show how parasitic gate effects can mimic nonreciprocity in topological insulators. We present methods to check whether nonreciprocity is gate-induced, either within a single device or by comparing a set of devices.

\section{Materials and methods}
VBST thin films are deposited by molecular beam epitaxy. We use the deposition protocol for \bst (BST) thin films \cite{mulder2022revisiting, Mulder_Glind_2023}, and include magnetic doping by evaporating vanadium from a Knudsen cell heated to $T = 1600$\degree C during deposition. The nominal thickness is 10 nm, with estimated $x \approx 0.72$ and $y \approx 0.01$ \cite{wielens2021axion}. We also fabricate nonmagnetic BST ($y = 0$) devices to exclude breaking time-reversal symmetry by intrinsic magnetization.
\begin{figure*}
    \centering
    \includegraphics[width=\textwidth]{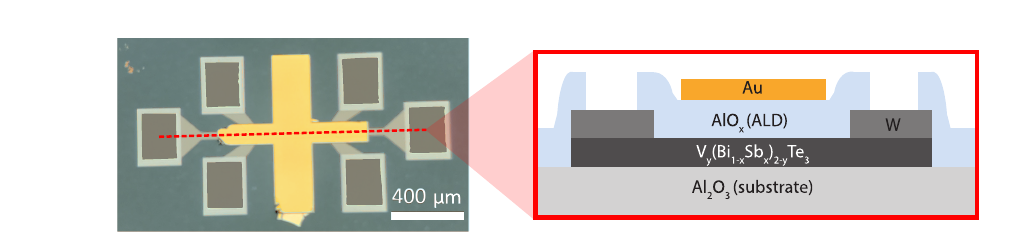}
    \caption{False-colored optical microscopy picture and schematic cross-section of fabricated devices (not to scale). The VBST film (thickness $t = 10$ nm) is deposited on an Al$_2$O$_3$ substrate and Ohmic contacts consist of sputter-deposited W ($t = 25$ nm). The entire device is capped with ALD-deposited AlO$_x$ ($t = 30$ nm), which forms the gate dielectric. The Au top gate electrode ($t = 75$ nm) covers the part of the Hall bar between, and including, both sets of Hall probes.}
    \label{fig:setup}
\end{figure*}

Figure~\ref{fig:setup} gives an overview of the device structure. The Hall bars are patterned using optical lithography and argon milling, and Ohmic contacts are created by sputter deposition of tungsten, with a thickness of 25 nm. The length and width between voltage probes is made to vary from device to device. The top gate dielectric is fabricated by atomic layer deposition (ALD) at $T = 100$ $^{\circ}$C and consists of 30 nm AlO$_x$. The oxide is covered by an Au gate contact.

Transport measurements are performed at 2.5 K in a physical properties measurement system (PPMS), where a DC bias current is sourced through the current lead by a Keithley 2401 Sourcemeter. 
If used, the top gate voltage ($V_\mathrm{TG}$) is applied relative to the source of the current lead. The longitudinal ($V_{xx}$) and Hall ($V_{xy}$) voltages are measured in a four-probe configuration, to omit contact and line resistances.

\section{Results and discussion}

\subsection{Gate-induced Hall voltage gradient}
We discuss a model describing the interplay between source-drain and top gate voltages. Two major simplifications should be noted: firstly, we consider a single topological surface state with top gate. The model can be expanded by including the capacitively coupled bottom surface \cite{fatemi2014electrostatic}, however, for the purpose of this analysis the single surface captures the physics. Secondly, the posed system of equations should be solved self-consistently to fit the model to the data. We only discuss the equations to first order, to extract scaling laws.

\begin{figure*}
    \centering
    \includegraphics{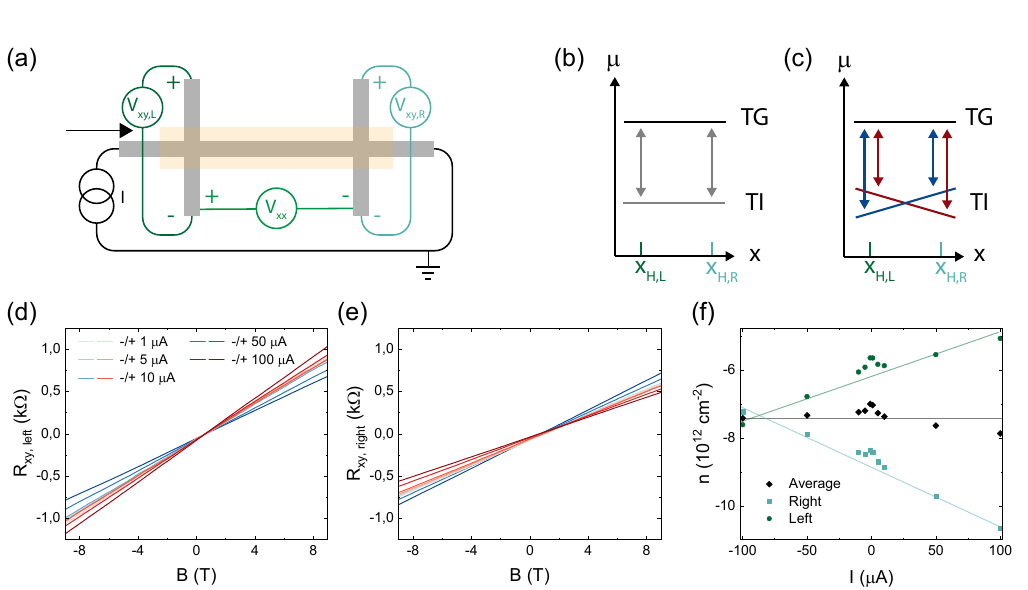}
    \caption{Gate-induced effects on the Hall voltage in a top-gated (Bi$_{1-x}$Sb$_{x}$)$_2$Te$_3$ Hall bar device at $T = 2.5$ K. The area covered by the top gate contact is highlighted in orange. (a) Measurement setup using multiple Hall probes. (b) For low bias currents, the chemical potential gradient along the topological insulator is negligible, and the potential difference between top gate (TG) and topological insulator (TI) is expected to be position-independent. (c) For high bias currents, a potential gradient along the device length can develop. The position-dependent Hall voltage is nonreciprocal in applied bias. The average potential difference between topological insulator and gate is unchanged. (d) Measured $R_{xy,L}$ and (e) $R_{xy,R}$ for varying negative (blue) and positive (red) DC bias currents. (f) Sheet carrier density extracted from (d) and (e) using a single-band Drude model, and the average of both carrier densities. The lines provide guides to the eye.}
    \label{fig:BST}
\end{figure*}

The devices are structured into Hall bars, see Fig.~\ref{fig:setup} and Fig.~\ref{fig:BST}(a). When a finite source drain bias is applied and the sample does not show ballistic conduction (i.e. does not show the quantum spin/anomalous Hall effect), we can assume that the chemical potential in the topological surface state has a gradient along the current lead. We assume that the resistance of the top gate contact $R_\mathrm{TG}\ll R_{xx}$, so we can treat the gate electrode as a single potential value. For small source-drain bias, the potential difference between the topological surface state and top gate is approximately constant, see Fig.~\ref{fig:BST}(b). For large source-drain bias in Fig.~\ref{fig:BST}(c), the gradient in topological surface state chemical potential locally changes the effective gate potential. The measured Hall effect thereby becomes dependent on the Hall probe position. In the following analysis, we find how these corrections scale with the applied current. In this section we consider the surface states of non-magnetic topological insulators, before generalizing to magnetic topological insulators in section~\ref{sec:gate_magTI}.

The effect of a top gate is to induce charges in a capacitively coupled device. The total charge induced by top gating ($n_\mathrm{TG}$) equals
\begin{equation}
    n_\mathrm{TG} = \frac{C_\mathrm{TG}}{e^2}\left[\mu_\mathrm{TG} - \mu(x) \right],
\end{equation}
with $C_\mathrm{TG}$ being top gate capacitance, $\mu_\mathrm{TG}$ the top gate chemical potential and $\mu(x)$ the chemical potential of the topological insulator surface as function of the position along the current lead. For our analysis, we are not searching for the total value of $n_\mathrm{TG}$, but for the \textit{additional} small change $\Delta n$ caused by the chemical potential gradient. As a DC bias current ($I$) changes $\mu(x) \propto I$, we will approximate $\Delta n(I, x) = n_\mathrm{TG}(I, x) - n_\mathrm{0} \approx c(x)I$ where $c$ is a proportionality factor depending on the spatial and current dependence of $\mu$ and the dielectric properties of the top gate oxide.

In a single-band Drude model, a change in carrier density $\Delta n$ results in
\begin{equation}\label{eq:rxy}
    R_{xy}(x) = \frac{B}{e\left[n_0 + \Delta n(x)\right]},
\end{equation}
where $R_{xy} = V_{xy}/I$, $B$ is the external out-of-plane magnetic field and $\Delta n$ is the change in carrier density with respect to equilibrium value $n_0$. 

Assuming $\Delta n \ll n_0$ (i.e. far away from the charge neutrality point \cite{footnote}), 
Eq. (\ref{eq:rxy}) becomes
\begin{equation}\label{eq:model_xy}
    R_{xy}(x) \approx \frac{B}{en_0}\left(1 - \frac{c(x)I}{n_0}\right).
\end{equation}
This equation implies a correction to the Hall voltage $\delta V_H \propto BI^2$, and $R_{xy}(I) - R_{xy}(-I) \propto I, B$. Furthermore, $R_{xy}$ becomes position-dependent, so the measured Hall effect depends on the choice of voltage probe location. 

We study the gate-induced Hall effect gradient in a (Bi$_{1-x}$Sb$_{x}$)$_{2}$Te$_3$ Hall bar with top gate contact, using the setup in Fig.~\ref{fig:BST}(a). First, we leave out vanadium-doping, to exclude breaking time-reversal symmetry due to magnetic dopants. In absence of ferromagnetism, the spin texture in topological insulators has been shown to lead to nonreciprocal effects \cite{he2018bilinear}. However, this effect is not position-dependent, in contrast with our gate-dependent model. During the measurements, the top gate contact is floating: it is not directly contacted by the measurement setup.

The Hall resistances measured at different locations along the current lead are shown in Fig.~\ref{fig:BST} (d) and (e). It is clear that the effect of a large DC bias current is opposite on both probe locations, in line with Fig.~\ref{fig:BST}(c). 

In Fig.~\ref{fig:BST}(f) we extract the carrier density from the Hall resistance using Eq. (\ref{eq:rxy}). 
The carrier densities on the left and right probes, and the average of both, are shown in  Fig. \ref{fig:BST}(f). Notably, there is an intrinsic inhomogeneity in the device, as the carrier densities on the left and right probes are not equal at low bias. Nevertheless, the carrier density changes linearly with DC current, in line with the approximations used to derive Eq. (\ref{eq:model_xy}). 

Moreover, the dominant charge carriers are $n$-type. A positive bias current would lower the effective gate potential on the left probe and raise it on the right. Therefore, a positive bias maximizes the $n$-type carriers on the right probe and minimizes the $n$-type carriers on the left probe (vice versa for negative bias), corresponding to the slopes in Fig. \ref{fig:BST}(f).

These results show that a position-dependent Hall effect is present in BST underneath a top gate contact. In the next section, we derive a nonreciprocal signal from Eq. (\ref{eq:model_xy}) and compare this to the measured data.

\subsection{From Hall voltage gradient to nonreciprocal resistance}
\begin{figure}
    \centering
    \includegraphics{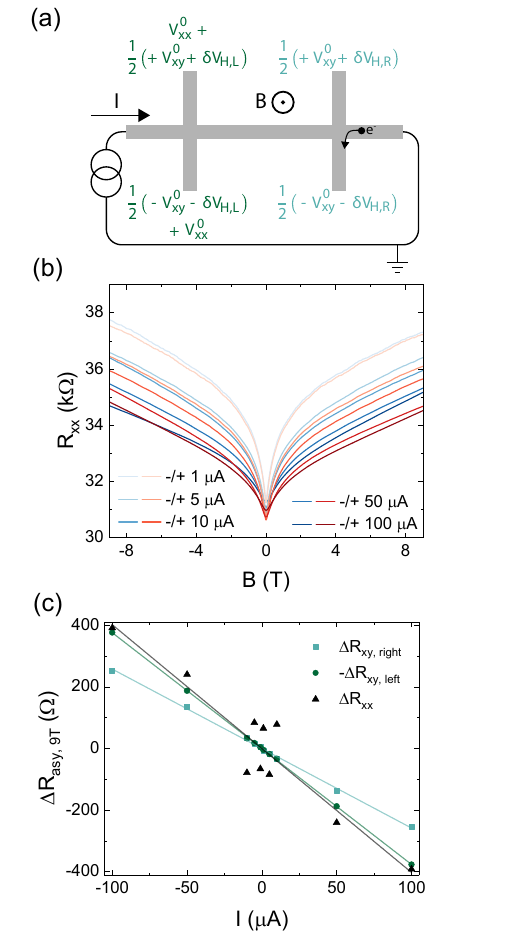}
    \caption{(a) Hall bar geometry with varying Hall voltage corrections along the current lead. $\delta V_\mathrm{H,L/R}$ cause a nonreciprocal contribution in $R_{xx}$ and $R_{xy,L/R}$ (Eq.~\ref{eq:nonrec}). (b) $R_{xx}$ corresponding to the dataset in Fig.~\ref{fig:BST}(d) and (e). A nonreciprocal component is visible, which grows with increased DC bias current and magnetic field. (d) $\Delta R_{xx}$ and $\Delta R_{xy,L/R}$ at $B = 9$ T after antisymmetrization in magnetic field, and linear fits through the data (following Eqs.~\ref{eq:model_xy} and \ref{eq:nonrec}).}
    \label{fig:BST_xx}
\end{figure}

In the previous section, we studied how a top gate contact can lead to a spatial gradient in Hall voltage, leading to Eq. (\ref{eq:model_xy}). We now derive the consequent nonreciprocal effects in detail. To start, we define the voltages in Fig.~\ref{fig:BST_xx}(a). These consist of the Hall voltage generated due to $n_0$ ($V_H^0$), the correction due to $\Delta n$ on the left/right probe ($\delta V_{H,L/R}$), and the longitudinal voltage.

Using the same measurement configuration as Fig.~\ref{fig:BST}(a), we write
\begin{equation}
\begin{aligned}
    &R_{xy,R} = R_{xy}^0 + \frac{\delta V_{H,R}}{I},\\
    &R_{xy,L} = R_{xy}^0 + \frac{\delta V_{H,L}}{I},\\
    &R_{xx} = R_{xx}^0 + \frac{1}{2}\left(-\frac{\delta V_{H,L}}{I} + \frac{\delta V_{H,R}}{I}\right).\\
\end{aligned}
\end{equation}
Here, $R_{xx}^0 = V_{xx}^0/I$ and $R_{xy}^0 = V_{xy}^0/I$ are the longitudinal and Hall resistances at low bias.

In order to calculate the nonreciprocal resistance, we need to answer the question: what happens if we reverse the current direction in Fig.~\ref{fig:BST_xx}?
Both $R_{xx}^0$ and $R_{xy}^0$ are unaffected by reversing the current direction ($V_{xy}$ reverses sign, but $I$ reverses sign as well, and the average carrier density setting the longitudinal resistance is unchanged). However, the \textit{corrections} to the Hall voltage will change, because the gradient in effective gate potential is mirrored across the vertical axis, see Fig.~\ref{fig:BST}(c). Therefore, if the film is homogeneous along the current lead, $\delta V_{H,L}(I>0) = -\delta V_{H,R}(I<0)$. 

Accordingly, the nonreciprocal resistances, given by $\Delta R = R(I) - R(-I)$, are
\begin{equation}\label{eq:nonrec}
\begin{aligned}
    &\Delta R_{xy,R} = -\frac{\delta V_{H,L}}{I} + \frac{\delta V_{H,R}}{I},\\
    &\Delta R_{xy,L} = \frac{\delta V_{H,L}}{I} - \frac{\delta V_{H,R}}{I} = -\Delta R_{xy,R},\\
    &\Delta R_{xx} = -\frac{\delta V_{H,L}}{I} + \frac{\delta V_{H,R}}{I} = \Delta R_{xy,R}.\\
\end{aligned}
\end{equation}
The nonreciprocal resistances in the above equations become nonzero when $\delta V_{H,L} \neq \delta V_{H,R}$, which is the case for the gate-induced Hall effect gradient. Using Eq. (\ref{eq:model_xy}), we find that the nonreciprocal resistance is antisymmetric in magnetic field, which is a useful tool in data analysis. 
One key result is the Hall voltage corrections emerging in $R_{xx}$ as well. This bears similarity with Eq. (\ref{eq:nonrec0}), because $\delta V_H \propto BI^2$.
Therefore, a nonreciprocal signal in a gated Hall bar device cannot be attributed directly to intrinsic chiral effects.

We now return to the experiments on the BST Hall bar with top gate contact. Figure~\ref{fig:BST_xx}(b) shows the longitudinal resistance for varying bias current. We observe the weak antilocalization cusp, typical for non-magnetic topological insulators \cite{lu2014weak, checkelsky2011bulk}. For increased bias currents the magnetoresistance is suppressed, possibly due to Joule heating \cite{nandi2018logarithmic}.
At high current and magnetic field, an asymmetry between positive and negative bias arises. At low magnetic field the asymmetry is less prominent, which ties into the average carrier density being constant in Fig.~\ref{fig:BST}(f) (the longitudinal resistance is an average quantity over the length of the current lead).

To test Eq. (\ref{eq:nonrec}), we antisymmetrize $R_{xx}$ and $R_{xy, L/R}$ in magnetic field and plot the nonreciprocal resistance at $B = 9$ T as function of bias current in Fig.~\ref{fig:BST_xx}(c). The magnitude of the nonreciprocal signals are of the same order, and the signs match Eq.~\ref{eq:nonrec}. Two remarks should be noted. Firstly, $\Delta R_{xy,L/R}$ slightly differ in magnitude, possibly due to inhomogeneity in the film which is not taken into account in the model. Secondly, $\Delta R_{xx}$ (after antisymmetrizing $R_{xx}$ in magnetic field) does not cross zero for small biases, which is possibly due to trapped charges in the (untrained) gate oxide adding a time-dependent drift to the signal. At low bias currents, this drift exceeds the nonreciprocal signal expected from Eq.~(\ref{eq:nonrec}). As the drift is absent in $\Delta R_{xy}$ we do not attribute it to a time-dependent change in carrier density, but it could result from leakage current to the gate electrode.

Concluding this section, the interplay between the potential gradient on the topological insulator surface arising at large bias currents, and the constant potential in the top gate electrode, can cause a nonreciprocal effect. This nonreciprocity affects both $R_{xx}$ and $R_{xy}$, which we observe in the nonmagnetic topological insulator BST. 

\subsection{Magnetic topological insulator}\label{sec:gate_magTI}
\begin{figure*}
    \centering
    \includegraphics[width=0.99\textwidth]{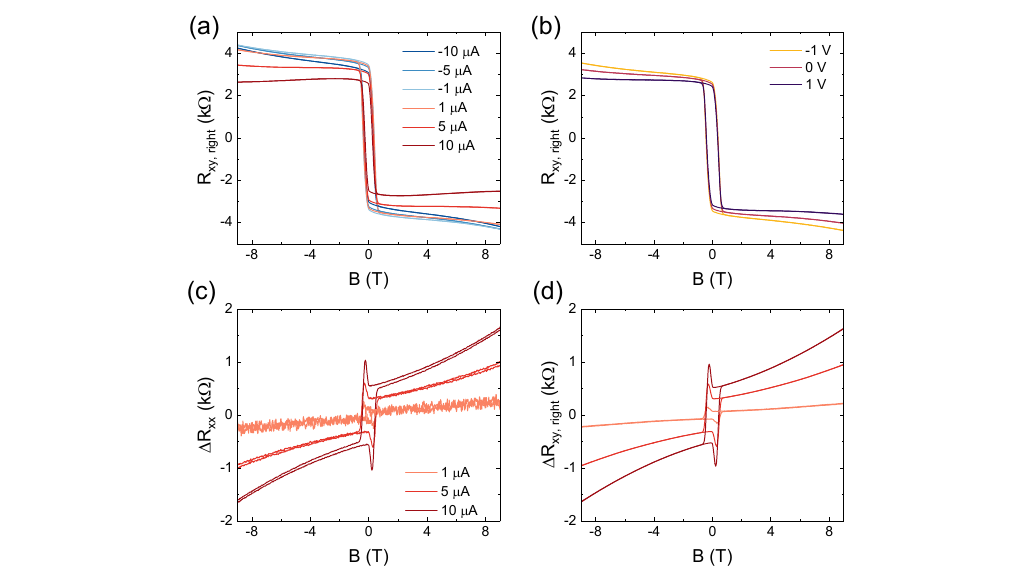}
    \caption{$R_{xy}$ in a V$_y$(Bi$_{1-x}$Sb$_{x}$)$_{2-y}$Te$_3$ Hall bar at $T = 2.5$ K. The setup resembles Fig.~\ref{fig:BST}(a) where $R_{xx}$ is measured between the bottom voltage probes and $R_{xy}$ at the right-most Hall probes. (a) $R_{xy}$ for varying DC bias current with a floating top gate contact. (b) $R_{xy}$ for varying top gate voltages, measured using standard lock-in techniques, with $I_\mathrm{RMS}$ = 100 nA. (c) $\Delta R_{xx}$ and (d) $\Delta R_{xy, R}$, both antisymmetrized in magnetic field. The nonreciprocal signals are equal, in line with Eq. (\ref{eq:nonrec}).}
    \label{fig:MR_VBST}
\end{figure*}
Having found gate-induced nonreciprocal effects in BST, we now turn toward the magnetically doped VBST. In similar materials (using Cr- instead of V-doping), a large nonreciprocal resistance has been observed \cite{yasuda2020large}. With the obtained knowledge of gate-induced nonreciprocity, it will be of interest to study whether nonreciprocal signals in our VBST are caused by gate-induced or intrinsic effects.

We model the anomalous Hall effect in VBST films using \cite{nagaosa2010anomalous}
\begin{equation}\label{eq:AHE}
    R_{xy} = R_\mathrm{AH} + R_\mathrm{OH},
\end{equation}
where $R_\mathrm{AH}$ and $R_\mathrm{OH}$ are the anomalous and ordinary Hall resistances, respectively. Assuming that the gate effect does not effect the sample magnetization, only the ordinary Hall effect will be influenced by gating. Consequently, Eq. (\ref{eq:nonrec}) holds for VBST as well. 

Figure~\ref{fig:MR_VBST}(a) shows the Hall resistance for a range of bias currents at $T = 2.5$ K, where a clear nonreciprocity is present. Note that we measure $R_{xy}$ on the rightmost set of Hall probes. Using the same device, we obtain the results in Fig.~\ref{fig:MR_VBST}(b) by varying the gate voltage instead of the bias current, measured using lock-in techniques with a 100 nA AC bias current at 13 Hz. Beyond the magnetic hysteresis, the sign of the Hall slope corresponds to a $p$-type doping in our measurement configuration. The device is not in the quantum anomalous Hall regime, where we would expect $R_{xx}\rightarrow 0$ and $R_{xy} \rightarrow h/e^2$ beyond the magnetic hysteresis \cite{tokura2018nonreciprocal, chang2013experimental}. Furthermore, we note that the coercive field is not influenced by changing the gate voltage. 

In this device $R_{xx} \sim 10^4$ $\Omega$ at 2.5 K, so $I = 100\ \mu$A would lead to $V_{xx}\sim 1$ V in Fig~\ref{fig:MR_VBST}(a). This order of magnitude is comparable to the range over which the gate voltage was varied, so the asymmetry arising from gate-induced effects is realistic. 

Next, we check whether the nonreciprocity follows Eq.~\ref{eq:nonrec}. We calculate the nonreciprocal $\Delta R_{xx}$ and $\Delta R_{xy, R}$ in Fig.~\ref{fig:MR_VBST}(c) and (d), and find that both are equal in magnitude. This matches Eq. (\ref{eq:nonrec}).

\begin{figure}
    \centering
    \includegraphics{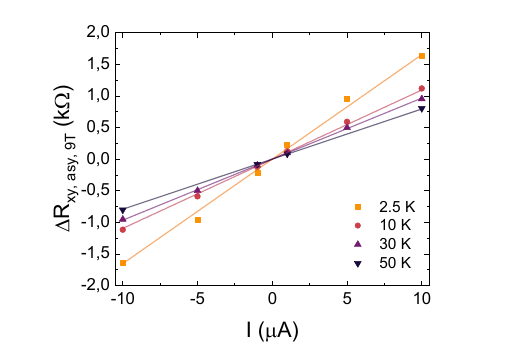}
    \caption{Nonreciprocity in the V$_y$(Bi$_{1-x}$Sb$_{x}$)$_{2-y}$Te$_3$ Hall bar of Fig.~\ref{fig:MR_VBST}. $\Delta R_{xy} = R_{xy}(10\ \mu\mathrm{A}) - R_{xy}(-10\ \mu\mathrm{A})$ at $B = 9$ T is calculated for a range of temperatures, after antisymmetrizing the data in magnetic field. $\Delta R_{xy}$ is nonzero far above the Curie temperature ($T_c$ $\approx$ $10$ K).}
    \label{fig:VBST_asy}
\end{figure}

If the nonreciprocity would originate from Eq. (\ref{eq:nonrec0}), it would also be antisymmetric in $B$ and $I$, so the results discussed up to now do not necessarily pinpoint the origin of the effect. Thus, we turn to the temperature dependence of the measured nonreciprocity: in similar materials, the nonreciprocal effects are expected to disappear above the Curie temperature if they are closely related to the ferromagnetic state \cite{yasuda2016large}. In Fig.~\ref{fig:VBST_asy}, $\Delta R_{xy}(I)$ at $B = 9$ T is shown for a range of temperatures, and notably $\Delta R_{xy}$ exists far above the Curie temperature ($T_c \approx 10$ K in our VBST films, based on the temperature at which $R_{xy}$ at $B = 0$ T changes from zero to a finite value, an approach utilized in previous research \cite{chang2013thin}). Combining this with Fig.~\ref{fig:MR_VBST} the nonreciprocal effect measured here is likely gate-induced.  

Finally, we study two VBST devices fabricated from the same thin film: one with top gate contact, and one without top gate contact. The device without top gate contact has been capped with AlO$_x$ (thickness 30 nm) as well to maintain similarities between both devices. We measure magnetotransport at low bias current in both devices and find that the top-gated device can be tuned to resemble the device without top gate in magnetoresistance.

\begin{figure}
    \centering
    \includegraphics[width=0.99\columnwidth]{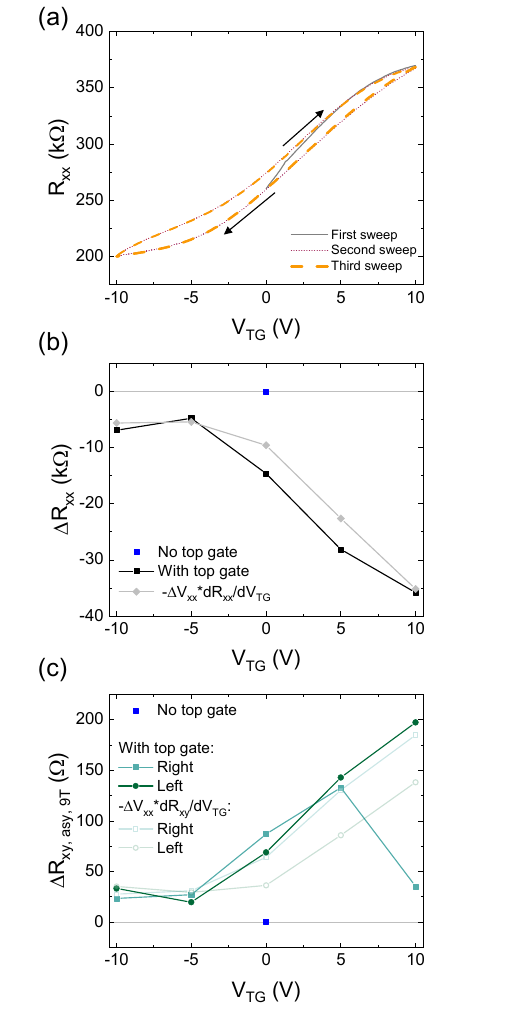}
    \caption{Comparison between two V$_y$(Bi$_{1-x}$Sb$_{x}$)$_{2-y}$Te$_3$ Hall bars, either with or without top gate, fabricated from the same thin film. All data is obtained at $T = 2.5$ K. (a) $R_{xx}$ as function of $V_\mathrm{TG}$ at $B = 0$ T during gate training of the top gated device, measured using a DC bias of $I = 1$ $\mu$A. (b) Nonreciprocal resistance $\Delta R_{xx}$ at $B = 0$ T in the Hall bar with(out) top gate (black/blue), compared with an estimate based on $R_{xx}(V_\mathrm{TG})$ following Eq.~(\ref{eq:approx_dR}) (grey). (c) Nonreciprocal resistance $\Delta R_{xy}$ at $B = 9$ T after antisymmetrization in magnetic field, on the left and right Hall probes. Lighter colored lines show the corresponding estimate from Eq.~\ref{eq:approx_dR}. $\Delta R_{xy}$ from the sample without top gate is shown in blue. In (b) and (c), $V_\mathrm{TG}$ was swept in the downward direction (10 V $\rightarrow$ -10 V).}
    \label{fig:gate_vs_nogate}
\end{figure}

The major difference of measurements on the top gated sample with the data discussed up to this point, is that here, the gate oxide has been trained by sweeping the gate voltage back-and-forth multiple times, until a reproducible (hysteretic) resistance as function of gate voltage was obtained, as shown in Fig.~\ref{fig:gate_vs_nogate}(a). The monotonic increase of $R_{xx}$ implies that the chemical potential of the topological insulator does not cross the Dirac point. Measurements of the nonreciprocal signal are performed in the same thermal cycle after gate training, while the gate voltage is regulated.

For the top-gated device, Fig.~\ref{fig:gate_vs_nogate}(b) shows a large nonreciprocal signal for all gate voltages, given by $\Delta R_{xx} = R_{xx}(I) - R_{xx}(-I)$, at $B = 0$ T. The signal being present without an external magnetic field could imply that the average carrier density in the film is affected by self-gating, when a DC bias current is applied. In this case, Eq. (\ref{eq:nonrec}) does not apply, because $R_{xx}^0$ gains a bias dependence through the changing average carrier density. The result is a different type of nonreciprocal signal: instead of $\Delta R_{xx}$ being set by corrections to the Hall voltage, the change in average carrier density will dominate the signal.

Because Eq. (\ref{eq:nonrec}) does not hold, we cannot use the same analysis as before. Alternatively, we can estimate the expected nonreciprocity from the slope of $R_{xx}(V_\mathrm{TG})$. If varying $V_\mathrm{TG}$ (at constant bias current) is similar to varying $V_{xx}$ by changing the bias current (at constant gate voltage), we can use a linearization of $R_{xx}(V_\mathrm{TG})$ to estimate 
\begin{equation}\label{eq:approx_dR}
    \Delta R_{xx} \approx -\dv{R_{xx}}{V_\mathrm{TG}} \Delta V_{xx},
\end{equation}
where $ \Delta V_{xx} = V_{xx}(I = 10\ \mu\mathrm{A}) - V_{xx}(I = -10\ \mu\mathrm{A})$. The results are shown in Fig.~\ref{fig:gate_vs_nogate}(b). In the gated device, Eq. (\ref{eq:approx_dR}) roughly follows the measured $\Delta R_{xx}$. In the un-gated device, $\Delta R_{xx}$ is negligible and does not fall within the range accessible by changing the top gate voltage.

Apart from $\Delta R_{xx}$, we check for signatures of the position-dependent Hall effect in $\Delta R_{xy}$ at $B = 9$ T in Fig.~\ref{fig:gate_vs_nogate}(c). In the gated device, $\Delta R_{xy}$ has the same sign on both Hall probes, showing again that Eq. (\ref{eq:nonrec}) does not hold. Both the datasets on left and right Hall probe locations match the estimate from Eq. (\ref{eq:approx_dR}) fairly well. Only the point at $V_\mathrm{TG} = 10$ V deviates, which might be an effect of hysteresis in the top gate training curve, as the measurements were performed in the downward sweep direction from 10 to -10 V. 

Most importantly, again the nonreciprocity is absent in the device without top gate.  
This result is key, showing that the presence of a top gate contact induces bias asymmetry in a Hall bar device, and the large nonreciprocity is not intrinsic to the VBST film.

\section{Conclusion}
In this manuscript we have shown that a top gate contact can cause nonreciprocity in Hall bar devices. Similar to previous experiments \cite{ye2022nonreciprocal, yasuda2020large}, $\Delta R$ depends linearly on bias current and is antisymmetric in magnetic field, making it a challenge to distinguish gate-induced effects from magnetochiral anisotropy intrinsic to the material. 

We measured nonreciprocal signals in both nonmagnetic (BST) and magnetic (VBST) topological insulator Hall bar devices. Comparing devices with and without a top gate is an effective method to identify the origin of nonreciprocity. However, such a comparison is not always available, as is the case for the VBST device in Fig.~\ref{fig:MR_VBST}.
Circling back to Eq. (\ref{eq:approx_dR}) and Fig.~\ref{fig:gate_vs_nogate}(b-c), we found that substituting changes in gate voltage for changes in the voltage across the topological insulator provides a reasonable approximation for $\Delta R$. 
Since the applied gate voltage in Fig.~\ref{fig:MR_VBST} is comparable to the voltage across the topological insulator, it is likely that the presence of the gate electrode caused the observed nonreciprocity in this device. This conclusion is strengthened by the nonreciprocity existing far above the Curie temperature.

We note that gate training effects have not been taken into account in this analysis. 
If the gate oxide leaks charge over time, the measured correspondence between bias and gate voltage can be altered.
Moreover, the effects of floating or fixing the gate potential are debatable, and subject to change after training the top gate oxide. 

The similarities between magnetochiral anisotropy and gate electrode-induced nonreciprocity vanish when expanding the range over which the devices are gate-tunable. 
We derived Eq. (\ref{eq:nonrec}) assuming that corrections to the local carrier density due to a gradient in effective gate potential are small compared to the intrinsic carrier density. The equations change to a different scaling regime when $\Delta n \sim n_0$ \cite{footnote}.  
On the other hand, if the VBST device can be gate-tuned through the charge-neutrality point, nonreciprocal effects related to interplay between edge and bulk states disappear \cite{yasuda2020large}.
Comparing the scaling of the nonreciprocal resistance over a gate voltage range therefore provides clarity on the origin of the effect.

To prevent parasitic gate electrode-induced nonreciprocity, the aspect ratio of Hall bar devices could be adapted such that the voltage resulting from the bias current will not be similar to the typical gate voltage range, taking into account design limits for proper Hall measurements \cite{ihn2009semiconductor}. On the other hand, using thicker gate oxides or back gates also prevents this. However, this reduces the available transport regime that can be probed within a set gate voltage range as well.

In summary, when searching for nonreciprocity in top-gated topological insulator devices, we recommend comparing measured resistance values as a function of bias and gate voltage. One method is measuring the Hall voltage at multiple probe locations. This way, a gradient in Hall voltage due to an effective gate voltage gradient can be uncovered, which can be related to nonreciprocal longitudinal and Hall voltages. Furthermore, the magnitude of the measured nonreciprocity can be related to the slope of the resistance as function of gate voltage. Finally, the strongest method is comparing whether nonreciprocity is present in similar devices with and without top gate, if possible. Any of these comparisons should be made before concluding the origin of nonreciprocal effects in top-gated topological insulator devices. 

\begin{acknowledgments}
We thank \.{I}nan\c{c} Adagideli and Sander Smink for fruitful discussions. 
Regarding MBE deposition, we would like to acknowledge Daan Wielens for valuable discussions on VBST optimization. 

S.K. and F.W. deposited VBST, fabricated devices. S.K. performed transport measurements and wrote the manuscript with input from all authors. A.B. supervised the project. This research was supported by a Lockheed Martin Corporation Research Grant.

\end{acknowledgments}

\bibliography{references}

\begin{thebibliography}{22}
\bibitem [{foo()}]{footnote}%
  \BibitemOpen
  \href@noop {} {}\bibinfo {note} {In contrast, around the Dirac point, $n_0 \approx 0 \ll \Delta n$. Then, $R_{xy} = B/(e\Delta n) \approx B/(ec(x)I)$.}\BibitemShut {Stop}%
\end{thebibliography}
\FloatBarrier
\clearpage

\widetext

\end{document}